\documentclass[a4paper,10pt]{article}
\usepackage[utf8]{inputenc}

\title{Differentially Private Synthetic Heavy-tailed Data}
\author{Tran Tran \and Matthew Reimherr \and Aleksandra Slavković}
\date{%
    \small The Pennsylvania State University, Department of Statistics \\
    Last Updated: October 13, 2023
}
\usepackage[letterpaper, portrait, margin=1in]{geometry}
\usepackage{natbib}
\usepackage{graphicx}
\usepackage{amsfonts}
\usepackage{amsmath}
\usepackage{amsthm}
\usepackage{float}
\usepackage[ruled,vlined]{algorithm2e}
\usepackage[super]{nth}
\usepackage[toc,page]{appendix}
\usepackage{diagbox}
\usepackage{caption}
\usepackage{subcaption}
\usepackage{adjustbox}
\usepackage{url}

\theoremstyle{definition}
\newtheorem{definition}{Definition}

\newcommand{\mbR}{\mathbb{R}}
\newcommand{\bx}{{\bf x}}

\begin{document}
\maketitle

\begin{abstract}
	 The U.S. Census Longitudinal Business Database (LBD) contains employment and payroll information of all U.S. establishments and firms dating back to 1976 and is an invaluable resource for economic research. However, the sensitive information in LBD requires confidentiality measures that the U.S. Census in part addressed by releasing a synthetic version (SynLBD) of the data to protect firms' privacy while ensuring data usability for research activities, but without provable privacy guarantees. Generating synthetic heavy-tailed data with a formal privacy guarantee while preserving high levels of utility is a challenging yet practically highly relevant problem for many data products not just LBD. In this paper, we evaluate using the framework of differential privacy (DP) that offers strong provable privacy protection against arbitrary adversaries to generate synthetic heavy-tailed data with a formal privacy guarantee while preserving high levels of data utility. We propose using the K-Norm Gradient Mechanism (KNG) with quantile regression for DP synthetic data generation. The proposed methodology offers the flexibility of the well-known exponential mechanism while adding less noise. We propose implementing KNG in a stepwise and sandwich manner such that new quantile estimation relies on previously sampled quantiles, to more efficiently use the privacy-loss budget.  Through a simulation study and application to the SynLBD, our results show that the proposed methods can achieve better data utility relative to the original KNG at the same privacy-loss budget. Importantly our DP-SynLBD can capture the economic trends such as gross employment often estimated from LBD, and point to directions that can lead to further improvements in estimation. 
\end{abstract}

\section{Introduction} \label{introduction}

Technological advances allow massive amounts of data to be collected from individuals and organizations and analyzed faster at a lower computational cost. This presents new and unique research opportunities for addressing many interesting problems. However, due to growing societal concerns about privacy attacks and loss of control over the collected data, especially those involving human subjects, a great amount of data is not publicly available and is only accessible through a secure platform. Such platforms are often monitored by data curators or disclosure boards or are publicly available after being modified via disclosure protection methods \citep{auxier2019americans, lee2021protecting, abowd20222020, census2023restricted}. The restricted access to data not only limits research progress in important human-related fields but also hinders reproducibility efforts in science. Developing reliable and efficient data privacy protection methods has received significant attention from society, governments, research organizations, and companies \citep{bailie2019abs, nayak2020metainitiative, bach2022differential, whitehouse2023privacyreports, apple2023dpscenes}.

One popular technique to mitigate the loss of privacy when releasing sensitive data is generating synthetic data \citep{little1993statistical, rubin1993statistical}. A \textit{synthetic dataset} is an artificially generated copy that shares the statistical properties of the original confidential dataset. Compared to traditional statistical disclosure control (SDC) methods, such as removing direct personal identification or data swapping, synthetic data offer more privacy protection by containing either no actual records (fully synthetic data) or some selected actual records or variables (partially synthetic data)\footnote{The terminology of full versus partial synthesis is typically interpreted as whether or not all of the variables in the dataset are synthesized. However, the real difference between the two approaches lies in the way that the synthetic observations are generated. For details see \citet{reiter2007multiple}.}. Since synthetic data have similar statistical properties to the original one, they also enable researchers to perform exploratory data analyses while waiting for approval to access the sensitive data. Researchers are actively expanding the literature on synthetic data, with topics ranging from the generation process \citep{raghunathan2003multiple, reiter2005using, reiter2009using, drechsler2011synthetic, raab2016practical} to statistical inferences \citep{raghunathan2003multiple, reiter2003inference, kinney2010tests, klein2015likelihood, slavkovic2010synthetic, pistner2018synthetic} for both fully and partially synthetic data. Many synthetic products have also been distributed publicly, such as the Synthetic Longitudinal Business Database (SynLBD) \citep{kinney2011towards} and the Survey of Income and Program Participation (SIPP) Synthetic Beta \citep{benedetto2013creation}.

However, synthetic data are not without fault. Many researchers have raised concerns about the usability and trustworthiness of the analysis of synthetic data \citep{woo2009global}. As a result, it is crucial to have standard guidelines to evaluate the quality of a synthetic dataset before releasing it for public use. Some notable works in this area include \citet{karr2006framework, woo2009global, drechsler2009disclosure}. Among them, \citet{snoke2018general} developed a utility score to assess the closeness in distribution as well as the similarity in specific statistical analysis results between the original data and the synthetic data. \citet{arnold2020really} also compiled a framework to evaluate synthetic data quality based on users' needs and perspectives.

One data type that commonly appears in sensitive datasets and plays a vital role in research activities across many fields is skewed data, such as income. However, synthesizing heavy-tailed data generation poses a dilemma: It is essential to both preserve the tail characteristics and protect the privacy of those isolated records. As a result, many researchers have introduced methods and techniques to mitigate this issue. \citet{hu2020risk} introduced a risk-weighted Bayesian approach, with individuals in the tails contributing less to the synthetic data. However, this method can lead to the loss of information in the tails and the reduction in data utility in the tails of the distribution. \citet{huckett2007microdata} and \cite{pistner2018synthetic} proposed generating synthetic data via quantile regression, as its flexibility and distribution-free property can help preserve the tails of the original data. This method results in synthetic data with a high level of utility but suffers from a lack of formal privacy guarantees.  

Differential privacy (DP) \citep{dwork2006calibrating} is the leading framework for statistical data privacy with formal mathematical guarantees. DP methods allow researchers and data analysts to carry out statistical inference and apply machine learning tools while limiting the amount of information one can learn about any specific subject. As DP provides a mathematically sound guarantee against disclosure risks and adversary attacks, this framework is being employed in government agencies (e.g., U.S. Census OnTheMap product in 2008 \citep{machanavajjhala2008privacy}, the 2020 U.S. Census \citep{abowd2018us, abowd20222020, abowd2023confidentiality}) and by many leading technology companies such as Apple \citep{apple2017dp, bittau2017prochlo}, Google \citep{erlingsson2014rappor}, Facebook \citep{evans2023statistically}, Microsoft \citep{ding2017collecting}, and Uber \citep{johnson2018towards}. DP synthetic data is considered a leading desirable solution to supported increased access to more sensitive data and has received much attention from researchers (\citet{li2014dpsynthesizer, snoke2018general, bowen2019comparative, Torkzadehmahani_2019_CVPR_Workshops, quick2021generating, mckenna2021winning, mckenna2022aim}, to name a few). 

The Longitudinal Business Database (LBD) is a key data product by the U.S. Census Bureau that contains employment and payroll information of all U.S. establishments and firms starting in 1976. Since it can provide insights into the U.S. labor market and business growth, the LBD is considered indispensable for research activities. However, the sensitive nature of the LBD requires rigorous safeguarding measures, thus restricting researchers' access to the data. Creating DP synthetic heavy-tailed data with a high level of utility is a potential solution to provide more data access. However, this task is challenging, due to the unique characteristics of skewed data and the difficulty in establishing a reasonable privacy-utility trade-off. 

To address this problem, we propose using the K-Norm Gradient mechanism (KNG, \cite{reimherr2019kng})  in the quantile regression setting to generate DP heavy-tailed synthetic data. KNG offers the flexibility of the well-known Exponential mechanism \citep{mcsherry2007mechanism} while adding a lower level of noise. However, applying KNG as-is would require independently fitting a large number of quantile regression models, each needing a fraction of the total privacy-loss budget, $\epsilon$. Even with a small privacy-loss budget per quantile, this approach can lead to a high total privacy-loss budget, especially when the data has many variables to synthesize. To address this concern, we modify the original mechanism and introduce the \textit{Stepwise KNG} and \textit{Sandwich KNG}. We derive the two extended mechanisms based on the solutions to the crossing quantile issue in the quantile regression literature. By estimating quantiles in a stepwise and sandwich order, we can incorporate the information given by previously estimated neighboring quantiles and lower the chance of an estimate going far off when the $\epsilon$ is small. We demonstrate that the two schemes can lead to more efficient use of the privacy budget and an increase in synthetic data utility. 

The structure of the paper is as follows: First, we provide a brief overview of the motivating application, quantile regression, differential privacy, the KNG mechanism, and synthetic data in Section \ref{backgrounds}. Second, in Section \ref{syn_kng}, we provide the algorithm to generate synthetic data through KNG and perform a simulation study to evaluate the performance of our proposed methods in Section \ref{syn}. Then, we illustrate the empirical application of this approach using SynLBD in Section \ref{app}. Finally, we conclude by discussing other areas of applications and potential improvements.

\section{Background} \label{backgrounds}

\subsection{Motivating Application}
The LBD contains information on all businesses with paid employees across all the industries in the U.S. \citep{lbdcensus2021}. LBD provides crucial measures of the dynamics of the economy and business activities. This dataset also helps researchers and policymakers understand and make informed decisions regarding the labor market, business growth trends, and market competition \citep{jarmin2002longitudinal}. LBD can either be used alone or in conjunction with other Census Bureau products, such as the Longitudinal Employer-Household Dynamics (LEHD) and the Management and Organizational Practices Survey (MOPS), to obtain further insights into the economy. It is the foundation for a wide range of micro and macroeconomic research studies, including underlying factors of job creation \citep{haltiwanger2013creates}, market volatility and unemployment \citep{davis2010business}, pay and productivity \citep{bloom2021pay}, and retail chains \citep{jarmin20096}. 

Additionally, LBD is the data source for the Business Dynamics Statistics (BDS) \citep{bdscensus2021}. BDS provides valuable information about labor market dynamics and business cycles, through records of job creation versus destruction rates as well as firm opening versus closing rates at both individual business and industry levels. As BDS is updated annually from 1976 to 2018, it plays a crucial role in economics, business strategy, and public policy research \citep{goetz2021business, goldschlag2020business, moscarini2012contribution, haltiwanger2009business}. 

Given the sensitive nature of the data collected, the U.S. Census Bureau strictly limits its access due to high-security544 risks for businesses. Because the dataset contains all firms in all U.S. industries, traditional SDC measures like data swapping do not guarantee enough protection against privacy attacks. As an alternative, the U.S. Census offers researchers access to the microdata via a synthetic version of LBD, SynLBD \citep{synlbd_data}. SynLBD is a simulated dataset and resembles its sensitive counterpart not at the observation level but at the distribution level. In addition, synthetic data facilitates research progress since access to SynLBD (previously via Cornell's Virtual Server \citep{synlbd_data} and currently through the Census Bureau \citep{census2023synlbd}) is less tedious and rigorous than access to the restricted confidential LBD.

Since its creation, SynLBD has gone through multiple updates with the aim of achieving a balance between data utility and disclosure risks \citep{kinney2011towards, kinney2014synlbd}. However, none of the updates include a formal privacy framework. This shortcoming is not unexpected, as allocating the privacy budget for a massive dataset of 20 million records is quite challenging. This task is even more complicated because we need to preserve the time-varying nature of firms and industries over 25 years, from 1976 to 2000. In addition, the data can be sparse in some industries or extremely skewed in others, complicating data modeling.

To the best of our knowledge, our work is the first one focused on making SynLBD differentially private. Our proposed formal privacy methods show promising results under the strongest privacy definition of $\epsilon$-DP and allowance for including the data of more recent years, which already comprises more than 8.5 million records in 2013 alone. Furthermore, this setup can be used as a benchmark for considering future directions such as relaxations of privacy definitions and improvements in data utility. 

\subsection{Quantile Regression} \label{quantile}
\citet{koenker1978regression} introduced quantile regression, an approach known for its distribution-free characteristics and the relaxation of assumptions typically required for linear models. The flexibility of quantile regression allows it to capture the behavior of extreme observations in a dataset. Quantile regression has a wide range of applications, ranging from economics and finance to ecology and epidemiology \citep{yu2003quantile}.

Let $\tau \in (0,1)$ and $Y\in \mathbb R$ be a random variable representing the outcome of interest. We say that $q_{\tau} \in \mbR$ is a $\tau$-quantile of $Y$ if it satisfies $P(Y < q_{\tau}) \le \tau$ and $P(Y>q_\tau) \le 1-\textbf{}\tau$. Using this notation, quantile regression assumes that the quantiles are linearly related to a vector of predictors, $X \in \mbR^{n\times p}$, as 

\begin{equation}
    q_{\tau} = X^\top \boldsymbol \beta_{\tau}.
\end{equation}
Given an iid sample $\{(y_i,\bx_i):i=1,\dots,n\}$ drawn from the distribution of $(Y,X)$, the estimation of coefficients $\boldsymbol \beta_{\tau} \in \mbR^{p\times 1}$ can be achieved through:
\begin{equation}
    \widehat{\boldsymbol \beta}_{\tau} = {\arg \min}_{\boldsymbol \beta} \sum_{i=1}^n \rho_{\tau}(y_i - \mathbf{x}_i^\top\boldsymbol \beta)
\end{equation}
where $\rho_{\tau}(x) = x(\tau - I\{x \le 0\})$ is known as the check loss.

There is an implicit assumption of monotonicity in quantile regression, which means that, for example, for a fixed $X$ the \nth{85} quantile should not exceed the \nth{90} quantile. Crossing quantiles happen when this assumption is violated for at least one pair of estimated quantiles.  In the absence of covariates, it is relatively easy to avoid this problem, but when covariates are involved the problem becomes much more delicate. Formally, we say there is a crossing quantile if, for some $\bx_i$, we have:
\begin{equation}
    \mathbf{x}_i^\top \widehat {\boldsymbol \beta}_{\tau} <  \mathbf{x}_i^\top \widehat {\boldsymbol \beta}_{\tau'}, \text{ for some } \tau > \tau'.
\end{equation}
Crossing quantiles in regression pose a problem as they can cause the response distribution to not be a proper probability distribution 
\citep{bondell2010noncrossing,liu2011simultaneous}. This issue can occur if quantile regression is performed for several quantiles individually. Proposals to address this include quantiles estimation either in a stepwise manner \citep{liu2009stepwise, muggeo2013estimating} or simultaneously \citep{bondell2010noncrossing, liu2011simultaneous, tokdar2012simultaneous}. In the case of linear quantile regression, \citet{koenker1984note} also proposed enforcing quantile curves to be parallel to avoid the crossing quantile phenomenon. In stark contrast to the non-private setting, we demonstrate how to take advantage of this phenomenon for private quantile regression by exploiting the fact that neighboring quantiles are likely to be similar; we add constraints to our sampling process to produce multiple quantile regression estimates at a dramatically reduced privacy budget.

\subsection{Differential Privacy and The K-Norm Gradient Mechanism} \label{dpkng}
\subsubsection{Differential Privacy} \label{dp}
Differential privacy (DP), introduced by \citet{dwork2006calibrating}, is a formal mathematical framework for privacy that allows information to be learned about the population while still preserving the privacy of individuals in the data. In particular, differential privacy provides probabilistic guarantees on disclosures, linkage attacks, and composition. It also makes group privacy feasible and is secure for post-processing. Notably, DP allows privacy loss quantification and enables privacy to go beyond a binary definition \citep{dwork2014algorithmic}.

Under DP, methods can be evaluated based on how efficient their privacy budget usage is, or in other words, how accurate they are with a fixed amount of privacy loss and vice versa. The classical DP framework quantifies the privacy loss of a mechanism through a parameter $\epsilon > 0$. The lower $\epsilon$ is, the more privacy protection a mechanism provides by adding a higher noise level to the statistical summaries that are to be publicly released. Typically, $\epsilon$ takes values less than 1, although in practice values of  $\epsilon$ of 5 or higher (e.g., \citet{abowd20222020, census2021parameters}) are considered and employed, even though they imply a higher level of privacy loss. 

Let $\mathcal{D}^n$ be a collection of databases with $n$ rows. $D, D' \in \mathcal{D}^n$ are considered adjacent databases if their Hamming distance, $\delta(D, D'):= \#\{i|D_i \neq D'_i\}$, is 1, i.e they differ in only a single record. Let $(\Theta, \mathcal{F})$ be measurable space equipped with a family of probability measures $\{\mu_D: D \in \mathcal{D}^n\}$ representing the privacy mechanism. Following the definition of \citet{dwork2006calibrating}, differential privacy can be formally defined as:

\begin{definition} \citep{dwork2006calibrating} 
A privacy mechanism $\{\mu_D: D \in \mathcal{D}^n\}$ satisfies $\epsilon$-differential privacy ($\epsilon$-DP) if for all $B \in \mathcal{F}$ and adjacent $D, D' \in \mathcal{D}^n$:
$$\mu_D(B) \le \mu_{D'}(B)\exp(\epsilon).$$
\end{definition}

\subsubsection{The K-Norm Gradient Mechanism (KNG)} \label{kng}
KNG \citep{reimherr2019kng} sanitizes a statistic by promoting solutions that minimize the gradient of the objective function. This mechanism behaves in a very similar way to the well-known exponential mechanism \citep{mcsherry2007mechanism}, but works directly with the gradient instead of the objective function itself. Taking advantage of the gradient makes KNG stand out among other private mechanisms: it does not limit the type of norms one can use. KNG also results in a more optimal amount of noise added to the summaries than the exponential mechanism. Compared to the level of statistical estimation error, KNG adds an asymptotically negligible amount of noise, while the exponential mechanism does not \citep{awan2019benefits}. 

Additionally, KNG has an advantage over objective perturbation, as it can achieve DP even when the objective function is not strongly convex and twice differentiable. This property makes KNG a highly practical mechanism applicable to a wide range of objective functions. The authors also point out that KNG reduces the complexities in calculation and computational efforts. Suppose the objective function is twice differentiable and convex. In that case, KNG is asymptotically an instantiation of the K-norm mechanism \citep{awan2021structure}, ensuring that the noise introduced is approximately the lowest possible amount to satisfy DP.

KNG applies to quantile regression, whose objective function is not strongly convex and twice differentiable, due to its generalizability property. Formally, \citet{reimherr2019kng} derived the sampling function of KNG in the quantile regression setting as:

\begin{equation}
    f_n(\theta) \propto \exp\Biggl\{ \frac{-\epsilon n}{4(1-\tau)C_X} \Biggl\| -\tau \frac{1}{n} \sum_{i = 1}^n x_i + \frac{1}{n} \sum_{y_i \le x_i^\top \theta} x_i \Biggr\| \Biggr\}.
    \label{eq: kngqr}
\end{equation}

The sensitivity $\Delta$ is $2(1-\tau)C_X$, where $\sup_{x_1}\|x_1\| \le C_X$. In this case, the sensitivity calculation under KNG only involves the bounds on X, making it significantly simpler than the exponential mechanism. However, a disadvantage of KNG when applied to quantile regression is that the theoretical asymptotic utility guarantee, which states that KNG results in $O_p\left(\frac{1}{n}\right)$-level noise, does not hold. Despite this drawback, the authors were able to show through a simulation study that the level of noise added is still asymptotically negligible and is much lower than that of the exponential mechanism. Additionally, note that in order for Equation \ref{eq: kngqr} to be integrable, we need to define it only over a bounded subset of the domain. Finally, we assume that $\theta$ has a normal base measure and c = 0.00001 for the rest of the paper. As a result, we sample from:
\begin{equation}
    f_n(\theta) \propto \exp\Biggl\{ \frac{-\epsilon n}{4(1-\tau)C_X} \Biggl\| -\tau \frac{1}{n} \sum_{i = 1}^n x_i + \frac{1}{n} \sum_{y_i \le x_i^\top \theta} x_i \Biggr\| - c\|\theta\|^2 \Biggr\}.
    \label{eq: kngqr_normal}
\end{equation}

\subsection{Synthetic Data} \label{syn}
\subsubsection{Generation Process} \label{gen_syn}
For synthetic data to be useful for research activities, it needs to preserve both the marginal and joint distributions of all the variables in the data. Otherwise, any exploratory data analysis and preliminary statistical modeling performed by researchers while waiting for access to the real data are rendered useless. As a result, the synthesis procedure typically involves sequential modeling. For example, to create a synthetic dataset with $p$ variables, one usually starts with $x_1$, then moves to synthesize $x_2|x_1$, $x_3|x_1, x_2$, and so on. The simulation and SynLBD application studies in Section \ref{sim} and \ref{app} are both fully synthesized datasets. Thus, we generate the synthetic version of the first sensitive variable $x_1$ by using only the intercept term. After that, we follow the sequential method mentioned above for the rest of the sensitive variables $x_2, ... x_p$, and then use the estimated coefficients and the previously synthesized variables to create the synthetic version of $x_2, ... x_p$. Although we do not demonstrate any examples with the partial synthesis in this paper, we should note that KNG is also applicable in that case.

When synthesizing variables with quantile regression, \citet{huckett2007microdata} proposed constructing a model on a set of quantiles, $0.001, 0.01, 0.02,..., 0.98, 0.99, 0.999$. The authors suggest sampling from a standard uniform distribution for each observation to find the quantile of interest and perform interpolation to obtain the predicted value corresponding to this quantile. In \citet{pistner2018synthetic}, the authors also first built a model using $b$ bins based on the interval $(0, 1)$. However, in their proposed algorithm they recommend randomly selecting a bin for each observation and generating a synthetic data value with the midpoint of that bin. The latter algorithm has two advantages over the former. First, it takes advantage of the theory behind inverse transform sampling \citep{pistner2018synthetic}. Second, it helps avoid computational costs from interpolation, especially in the case of complicated models.

We follow \citet{pistner2018synthetic} for our synthetic data generation procedure. However, we replace quantile regression with differentially private quantile regression fitted via KNG and its extended methods as the synthesis model (Step 1a, Algorithm \ref{alg: syndata}).

\subsubsection{Utility and Other Performance Measures}
Synthetic data need to have the right balance between privacy and usability. If a synthetic dataset contains too much noise, it will fail to capture the characteristics of the original dataset, becoming meaningless for research and policy implications. On the other hand, if synthetic data resemble confidential ones too closely, they will be helpful for researchers but might lose their privacy guarantee. Due to this trade-off, the topic of synthetic data performance measures has received much attention from researchers (e.g., \cite{karr2006framework, woo2009global}) and it is ongoing. 

In particular, \citet{snoke2018general} reviewed and proposed two broad categories of utility measures, general and specific,  to quantify how closely a synthetic dataset resembles the original one. Specific utility evaluates the quality of synthetic data through the similarity in statistical data analysis results obtained on the synthetic and original data. To quantify specific utility, they consider two metrics: the overlaps in confidence intervals and standardized differences in summary statistics. General utility involves measuring how close the distribution of the synthetic data is to that of the original data. The underlying idea of general utility is that the closer the two datasets are in distribution, the harder it is for classification models to distinguish which records belong to which dataset. The propensity score mean-squared error, \textit{pMSE}, ranges from 0 to 0.25. The closer the pMSE is to 0, the harder it is for a model to discriminate between the two datasets, and the better the utility of synthetic data. To make comparisons among different synthesis procedures consistent, the authors also presented the idea of standardizing the pMSE score by its null distribution. All measures introduced above are generalizable to both fully and partially synthesized datasets. Following \citet{snoke2018general}, we calculated the pMSE score from a logistic regression model with and without interaction effects to evaluate the performance of our methods. In cases where the classical logistic regression model does not converge due to the sparsity of the data, we opt to use the logistic regression model with Ridge and LASSO to obtain the pMSE score.

Additionally, \citet{arnold2020really} established a utility framework from the data users' point of view. According to the authors, data users want inference and prediction based on synthetic data to be accurate compared to the original data and generalizable to the unobserved population data. Thus, to thoroughly evaluate a synthesizer, they proposed splitting the sensitive data and using the training and holdout set as comparison baselines along two main dimensions. The first dimension involves examining distributional closeness between synthetic data and the original data used to generate them (\textit{training data similarity)} versus the underlying population (\textit{generalization similarity}). The second dimension concerns general and specific utility, similar to the ideas proposed by \citet{snoke2018general} above. More precisely, to measure general data quality using training data, \cite{arnold2020really} suggest using the pMSE and Wasserstein distance ratios. In both cases, ratios close to 0 indicate good data utility. Similarly, we can quantify how close the synthetic dataset is to the holdout dataset. To assess specific data quality with training data, one should compute the average percentage of bias and the ratio of synthetic data coefficients' variances and covariances to those of training data. Finally, to capture specific data quality when using holdout data as the standard, the authors suggest a few options, such as the average percent of bias between synthetic data parameters and the population ones, the average time the 90\% confidence intervals capture the actual underlying value, and the average out-of-sample prediction root mean squared error (RMSE). \citet{arnold2020really} also recommend averaging results across multiple DP synthetic datasets under different data synthesizers to obtain a reasonable utility estimate for a specific DP method.

\section{Synthetic Data via the KNG Mechanism} \label{syn_kng}
In this section, we present two algorithms based on KNG for producing DP quantiles, which in turn will help us produce DP synthetic data.  

Originally proposed KNG mechanism estimates each quantile separately, and thus it is prone to encounter crossing problems, such as having a \nth{35} quantile that is lower than an \nth{25} quantile, as discussed above. The lack of data at the upper quantiles makes it even more difficult to estimate individually fitted quantile curves without violating the implicit monotonicity of quantile regression. This issue can potentially lead to unstable estimations in the upper tails and consequently low synthetic data utility.

Furthermore, independently estimating each quantile is not the most efficient way to use the privacy budget. More specifically, we need to spend a fraction of the budget on each quantile we want to estimate. The lack of data and the amount of noise added may require more privacy budget to get reasonable estimates. In other words, a substantial utility budget may be required to achieve acceptable accuracy for a large dataset with many variables using a grid of quantiles. Therefore, it is necessary to find a solution to incorporate the information on characteristics of the data obtained through previously estimated quantiles, which are likely to be similar. Exploiting information from neighboring quantiles can help ensure more efficient privacy budget usage while still enhancing the stability and closeness to the truth of the synthetic data. 

To avoid crossing quantiles in the non-private quantile regression setting, \citet{koenker1984note} suggested imposing equality constraints on the slopes. Instead of estimating both intercepts and slopes for each of the $m$ quantiles, it is only necessary to find one slope and $m$ different intercepts. With $\hat{\beta}_1$ as the intercept and $\hat{\beta}_k$, $k \ge 2$ as the slopes, this constraint can be expressed as:
\begin{equation}
    \hat{\beta}_{1, \tau} < \hat{\beta}_{1, \tau_1} \text{ and } \hat{\beta}_{k, \tau} = \hat{\beta}_{k, \tau_1} \text{ for } \tau < \tau_1 \text{ and } k = 2, 3,...
    \label{eq: stepwiseKoenker}
\end{equation}

\citet{liu2009stepwise} introduced the approach of conducting the quantile regression in a stepwise scheme to avoid crossing problems. Within this procedure, one should estimate the median first and treat it as an anchor quantile due to its low variance and high accuracy \citep{liu2009stepwise, muggeo2013estimating}. After that, other quantiles are computed iteratively based on the previously found quantiles. When $\tau < \tau_1$, $\hat{\boldsymbol \beta}_{\tau}$ can be used as the floor constraints to find $\hat{\boldsymbol \beta}_{\tau_1}$. On the contrary, when $\tau > \tau_1$, $\hat{\boldsymbol \beta}_{\tau}$ will be used as the cap. Based on this idea, we propose implementing the following constraint within the Markov chain sampling process in the proposed KNG mechanism:

\begin{equation}
    \mathbf{X}^\top \hat{\boldsymbol \beta}_{\tau} < \mathbf{X}^\top \hat{\boldsymbol \beta}_{\tau_1} \text{ for } \tau < \tau_1
    \label{eq: stepwiseDataCheck}.
\end{equation}

In addition, we use the all-at-once Metropolis-Hastings (AMH) algorithm to sample from the KNG function. We notice that the coefficients are highly correlated and can converge better if we sample them altogether. To avoid the cases where the chains go off the possible value region when $\epsilon$ is extremely low, we introduce soft lower and upper bounds in the algorithm, which can be set based on prior knowledge of the dataset. For example, a dataset involving the number of people cannot have negative values, thus having 0 as the lower bound. Meanwhile, the upper bound can be set as an arbitrary number times the maximum number of people in the data as a preventive measure. We note that setting the bounds is not a requirement to implement our algorithm, but is a way of taking advantage of prior knowledge to enhance the synthetic data utility. If we do not want to use bounds in the algorithm, the lower and upper bounds can simply be set as $-\infty$ and $\infty$.  Combining the above results, we propose the Algorithm~\ref{alg: stepwiseKNG} to find differentially private quantiles:

\vspace*{6pt} 

\begin{algorithm}[H]
\SetAlgoLined
\KwIn{An $(n \times k)$ design matrix $\mathbf{X}$, an $(n \times 1)$ response matrix $\mathbf{Y}$, and a vector of $m$ quantiles $\boldsymbol\tau$}
\KwOut{An $(k \times m)$ matrix of coefficients $\boldsymbol \beta$}

1. Sample $ \hat{\boldsymbol \beta}_{0.5}$ from Eq. \ref{eq: kngqr} with AMH using the non-private estimates as starting values\;
2. Split $\boldsymbol \tau$ into a vector of lower quantiles $\boldsymbol \tau_1$ and a vector of upper quantiles $\boldsymbol \tau_2$\;
3. Sort $\boldsymbol \tau_1$ in a decreasing order and $\boldsymbol \tau_2$ in an increasing order\;

\For{each $\tau$ in $\boldsymbol \tau_1, \boldsymbol \tau_2$}{
    4. Sample $\hat{\boldsymbol \beta}_{\tau}$ from Eq. \ref{eq: kngqr}\;
    5. If the non-crossing constraint from Eq. \ref{eq: stepwiseKoenker} or Eq. \ref{eq: stepwiseDataCheck} is satisfied, accept. Else, reject\;
    6. Repeat step 4, 5, and 6\;
    7. Use the accepted $\hat{\boldsymbol \beta}_{\tau}$ as the constraint for the next quantile\;
}
 \caption{Stepwise KNG Algorithm}
 \label{alg: stepwiseKNG}
\end{algorithm}

\vspace*{6pt} 

In Algorithm \ref{alg: stepwiseKNG}, we build the coefficients of new quantiles upon the results of the \nth{50} quantile. Therefore, instead of equally allocating the privacy budget at each quantile, we suggest spending more privacy budget at the \nth{50} quantile and dividing the remaining budget equally for the other quantiles. 


Motivated to further guarantee data utility even at a low privacy budget, we propose the sandwich KNG algorithm. Our reasoning is that once the upper and lower bounds have been established, the marginal privacy budget needed to find the quantiles between the two bounds is minimal. To implement this method, we need to first estimate the so-called anchor quantiles, such as \nth{5}, \nth{25}, \nth{50}, \nth{75}, and \nth{95}, using the stepwise KNG algorithm. The non-crossing constraint from Equation \ref{eq: stepwiseKoenker} can be revised as:
 \begin{equation}
     \hat{\beta}_{1, \tau_1} < \hat{\beta}_{1, \tau} < \hat{\beta}_{1, \tau_2} \text{ and } \hat{\beta}_{k, \tau_1} = \hat{\beta}_{k, \tau} = \hat{\beta}_{k, \tau_2} \text{ for } \tau_1 < \tau < \tau_2 \text{ and } k \ge 2
     \label{eq: sandwichKoenker}.
 \end{equation}

Similarly, the constraint from Equation \ref{eq: stepwiseDataCheck} also needs to be modified as:
\begin{equation}
    \mathbf{X}^\top \hat{\boldsymbol \beta}_{\tau_1} < \mathbf{X}^\top \hat{\boldsymbol \beta}_{\tau} < \mathbf{X}^\top \hat{\boldsymbol \beta}_{\tau_2} \text{ for } \tau_1 < \tau < \tau_2
    \label{eq: sandwichDataCheck}.
\end{equation}

The details of the sandwich KNG algorithm are as follows:

\vspace*{6pt} 
\begin{algorithm}[H]
\SetAlgoLined
\KwIn{An $(n \times k)$ design matrix $\mathbf{X}$, an $(n \times 1)$ response matrix $\mathbf{Y}$, a vector of $m$ quantiles $\boldsymbol\tau$, and a vector of $m'$ anchor/main quantiles $\boldsymbol\tau'$}
\KwOut{An $(k \times m)$ matrix of coefficients $\boldsymbol \beta$}
1. Estimate $\hat{\boldsymbol \beta}_{0.5}$ using Eq. \ref{eq: kngqr} with AMH using non-private estimates as starting values\;
2. Use Algorithm \ref{alg: stepwiseKNG} to compute $\hat{\boldsymbol\beta}_{\tau}$ for each $\tau$ in $\boldsymbol\tau'$\;
3. Consolidate the remaining quantiles into one vector $\boldsymbol \tau''$\;

\For{each $\tau$ in $\boldsymbol \tau''$} {
    4. Find the next lower and higher estimated quantiles, $\hat{\boldsymbol \beta}_{\tau_1}$ and $\hat{\boldsymbol \beta}_{\tau_2}$\;
    5. Sample $\hat{\boldsymbol \beta}_{\tau}$ from Eq. \ref{eq: kngqr}\; 
    6. If the non-crossing constraint from Eq. \ref{eq: sandwichKoenker} or Eq. \ref{eq: sandwichDataCheck} is satisfied, accept. Else, reject.\;
    7. Repeat step 5-6\;
    8. Update the list of estimated quantiles \;
}

 \caption{Sandwich KNG Algorithm}
 \label{alg: swKNG}
\end{algorithm}
\vspace*{6pt} 

\section{Simulation Study} \label{sim}

To provide a comprehensive evaluation of our proposed methods, in addition to the non-private synthetic data generated via quantile regression, we also compare our results to another $\epsilon$-DP method, the pMSE mechanism \citep{snoke2018pmse}. The pMSE mechanism tunes and releases the synthetic data version with the maximal distributional similarity. The pMSE mechanism optimizes the synthetic data output with respect to the pMSE score, while our methods do not have an integrated evaluating metric. Through simulation, \citet{snoke2018pmse} showed that under normally distributed data, the pMSE mechanism performs at least as well as other standard DP mechanisms, such as smooth histogram \citep{wasserman2010statistical} and noisy Bayes \citep{bowen2020comparative}, across different levels of privacy budget. Here, we make a comparison in the setting of skewed data and using additional utility measures.

\subsection{Simulation Set Up}

To compare the performance of the original, stepwise, and sandwich KNG mechanisms, we conduct a simulation study using 100 training and 100 testing datasets, each with a sample size of $n = 5000$. Each pair of training and testing datasets is generated under a different seed to ensure generalizability. We use the training datasets to generate synthetic data while keeping the testing data for data utility evaluation. We set $X_1 \sim Exp(0.1)$, $X_2|X_1 = 4 + 3X_1 + \xi_i$, $X_3|X_1, X_2 = 3 + 2X_1 + X_2 + \gamma_i$, where $\xi_i, \gamma_i \sim Exp(0.1)$. 

We allocate a total privacy budget of $1$ for the whole dataset. Specifically, we allocate a budget of 0.5 for $X_1$ as it is the first variable to be synthesized and can affect the overall data utility while dividing the remaining 0.5 equally between $X_2$ and $X_3$. To generate the synthetic version of each variable, we use a set of 49 quantiles (\nth{1}, \nth{3}, ..., \nth{47}, \nth{50}, \nth{53}, ..., \nth{97}, \nth{99}), and 6 main quantiles (\nth{5}, \nth{25}, \nth{50}, \nth{75}, \nth{95}, \nth{99}). Because each quantile is computed individually in the original KNG mechanism, we divide the privacy budget equally for each quantile. For the stepwise KNG mechanisms, we allocate 80\% of the budget available to the median and equally divide the rest among the remaining quantiles when synthesizing $X_2$ and $X_3$. Similarly, for sandwich KNG, we spend 80\% of the total budget on the main quantiles and 80\% of the main quantiles' privacy budget on the median. However, as we generate $X_1$ without any predictors, we assign 25\% of the total privacy budget to the median for stepwise KNG while allocating 60\% of the total privacy budget to the main quantiles, 25\% of which to the median for sandwich KNG. The privacy budget allocation can be adjusted based on the number of quantiles chosen for the procedure and the characteristics of the sensitive data. However, we generally recommend allocating more budget to the median and possibly the main quantiles. 


Additionally, as detailed in Section \ref{gen_syn}, we first synthesize $X_1$ through private quantiles, then use the synthetic $X_1$ to generate synthetic $X_2$, and finally, conditioning on synthetic $X_1$ and $X_2$ to find $X_3$. Our algorithm requires a bound on the predictors ($C_X)$ for the sensitivity calculation. Therefore, we assume a bound of 1, 46, and 106 when generating $X_1$, $X_2$, and $X_3$, respectively. The bounds when generating $X_2$ and $X_3$ are set based on the \nth{99} percentile of original predictors $X_1$ and $X_2$. The bound used when generating $X_1$ is set as 1 because we generate $X_1$ without any predictor. Any observations with values exceeding the set bounds are top coded, or winsorized, with the respective bound. In other words, if the upper bound is set as 46, and there are three observations exceeding 46, then they are replaced with 46. We decide to follow this strategy to mimic how bounds are set for sensitive data in practice \citep{crimi2014top}. For example, a data curator will use other public data sources to estimate an upper bound with a reasonable balance between capturing the heavy-tailed characteristic and ensuring the scale of noise added to the data is not too large.

We also assume upper and lower bounds on the predicted outcome and reject samples that fall out of this range, so that the Markov chains are more likely to explore the probable regions. For example, if we synthesize the number of employees of firms, we know that it cannot be negative and cannot be larger than the total number of employees in that industry. Note that the bounds are only meant to help guide the Markov chain and thus, do not need to be set strictly. For our simulation, we assume a lower bound of 0, and upper bounds of 1000 for $X_1$ and $X_2$, and 2000 for $X_3$.

\subsection{General Utility}
We use three utility measures, pMSE score \citep{snoke2018general}, Wasserstein randomization test \citep{arnold2020really}, and k-marginals \citep{task2021km} to assess the similarity in distribution between the raw and synthetic data. The idea behind the pMSE score is that if the synthetic version shares close statistical properties with the original one, the logistic regression model will not classify the observations of the two datasets well. After obtaining the synthetic dataset, we combine the sensitive and the synthetic datasets and train the logistic regression model to predict which observation belongs to which dataset. Let $\hat{p}_i$ be the predicted probability for observation $i^{th}$, $N$ be the total number of observations in the combined dataset, and $c$ be the proportion of synthetic data, then pMSE = $(1/N) \sum_{i=1}^N (\hat{p}_i - c)^2$. The pMSE score ranges from 0 to 0.25, and the lower the score is, the better. We test our methods using a regression model with interaction (denoted pMSE WI in Table \ref{tab: sim_general_utility}) and without interaction effect (pMSE WOI) between variables. As the model with interaction is more complex than the one without, we anticipate that it is more likely to distinguish between the two versions better, leading to a higher pMSE score. 

\citet{arnold2020really} proposed the Wasserstein randomization test (WRT) as a way to standardize data utility comparison using the Wasserstein metric. In order to conduct this test, we find the Wasserstein distance between two datasets after swapping a random observation between them and repeat this procedure a large number of times, such as 1,000 or 10,000. After that, we can approximate the null distribution of the distances by taking its median. Finally, we compute the ratio by dividing the Wasserstein distance between the original and synthetic datasets and the null distribution. A ratio of 0 means that the original version is exactly the same as the synthetic version, and higher scores indicate greater distributional differences. 

The k-marginal (KM) test, introduced by \citet{task2021km}, evaluates the utility of a synthetic dataset by estimating the difference in density of $k$ randomly selected variables between it and the original version. In particular, we create six bins based on the range of the data, ($\infty$, Min), (Min, Q1), (Q1, Median), (Median, Q3), (Q3, Max), and (Max, $-\infty)$, and calculate the difference in frequency of the data falling into a combination of bins of all three variables. For example, suppose 10\% of the real data falls into Bin 2 for all variables $X_1$, $X_2$, and $X_3$,  but only 6\% of the synthetic data belongs to these three bins. Then, the difference is 4\% or 0.04. Repeating this step for all 216 combinations of bins and summing them up to obtain the total density difference. This difference is plugged into a formula to convert it to an easier-to-interpret range, $\text{Score} = 1000 ((2 - \text{total difference})/2)$. The KM score ranges from 0 to 1000, where a score of 0 means complete difference and 1000 means complete similarity. In our simulation study, we use all three variables for this test.

For each method, we create a synthetic dataset based on a training dataset, using a privacy budget of $\epsilon = 1$ where applicable. We then repeat this procedure for 100 training datasets generated using 100 different seeds. The results in Table \ref{tab: sim_general_utility} are the mean and standard error of the utility of 100 synthetic datasets. By doing this, we aim to approximate the average performance of each method under $\epsilon = 1$.  

\begin{table}[H]
    \centering
    \begin{tabular}{ |c|c|c|c|c| } 
        \hline
        \diagbox{Method}{Utility} & pMSE WOI & pMSE WI & WRT & KM \\ 
        \hline
        Non-private         & 0.0003 (0.00001)  & 0.0004 (0.00001)&  2.86  (0.07) & 952.34 (0.67) \\
        Stepwise - Fixed    & 0.0146 (0.0004)   & 0.0167 (0.0004) &  12.65 (0.23) & 743.62 (3.29) \\ 
        Stepwise - Varying  & 0.0067 (0.0003)   & 0.0105 (0.0003) &  12.36 (0.22) & 531.40 (4.76) \\ 
        Sandwich - Fixed    & 0.0083 (0.0004)   & 0.0093 (0.0004) &  5.15  (0.12) & 762.32 (2.70) \\
        Sandwich - Varying  & 0.0077 (0.0004)   & 0.0084 (0.0004) &  4.88  (0.11) & 743.99 (4.95) \\
        KNG                 & 0.1219 (0.0018)   & 0.1309 (0.0024) &  42.18 (0.52) & 49.30  (1.83) \\
        pMSE Mechanism      & 0.0374 (0.0031)   & 0.0601 (0.0030) &  12.94 (0.85) & 742.81 (6.66) \\
        
        \hline
    \end{tabular}
    \caption{Comparing the general utility of synthetic simulated data generated by different methods. Means and standard errors (in parentheses) across 100 replications are reported.}
    \label{tab: sim_general_utility}
\end{table}

We find that adding structures to the way we estimate quantiles helps improve the general utility of the synthetic data. Our proposed methods perform better than KNG across all utility measures. We also observe that sandwich KNG schemes perform slightly better than their stepwise counterparts across all utility measures, but the differences between methods are not significant. This means that depending on the structure of the sensitive data, one of our methods may perform better than another.

Compared to the non-private version, our two best performers, sandwich fixed and varying slope KNG, have roughly 20\% lower KM scores, 1.75 times higher WRT scores, and 30 times higher pMSE scores under the regression model without and with interaction, respectively. These results indicate that we trade some distributional similarity between synthetic and original datasets for the $\epsilon$-DP protection. However, the general utility performance of our DP versions can be considered reasonably good, given that we only use a small privacy-loss budget of 1.

Our proposed methods perform either as well as or slightly better than the pMSE mechanism across all different general utility measures, except for the KM score for stepwise varying slope KNG. Since \citet{snoke2018pmse} pointed out that their mechanism may perform poorly if the distribution of data is misspecified, we present results based on the true distribution of the simulated data. This result shows that the proposed KNG mechanism with quantile regression on average adds less noise than the exponential mechanism, even when the latter has a built-in utility-maximizing function. This improvement in synthetic data utility can also be attributed to the fact that quantile regression is flexible and can address problems in modeling skewed data. On the other hand, the pMSE mechanism performs much better than the original KNG, showing that adding structures to the way we sample quantiles does pay off. We note that this is only the result under skewed data at $\epsilon = 1$. Therefore, it would be interesting to continue the performance evaluation under different data types and varying privacy budgets. 

Finally, to give a visual overview of the distributional similarity, we plot the distribution of synthetic data generated using different methods in Appendix \ref{appendix_simulation}. In general, thanks to the constraints in the sampling process, the synthetic versions generated using our proposed methods seem to be more likely to fall within the range of the actual data, compared to the original KNG. All the KNG methods, however, have more "bumpy" density plots than the pMSE mechanisms. We attribute this result to the fact that quantile regression is distribution-free, while we have to assume the data is exponentially distributed in the case of the pMSE mechanism.

\subsection{Specific Utility}
We evaluate the specific utility of synthetic data using the standardized coefficient difference and root mean squared error (RMSE) on testing data \citep{arnold2020really}. The standardized coefficient difference demonstrates how closely the regression model coefficients fitted using synthetic data resemble the results under sensitive data. We fit a standard linear regression model of $X_2$ based on $X_1$ on both the actual data and its synthetic version to obtain $\theta_j^{original}$ and $\theta_j^{synthetic}$, where $j = 0, 1$. Then the standardized coefficient difference can be calculated using the formula $\frac{|\theta_j^{original} - \theta_j^{synthetic}|}{SE(\theta_j^{original})}$. The smaller the difference is, the better the utility of the synthetic dataset.

Normalized RMSE (NRMSE), on the other hand, assesses the generalization and prediction ability of synthetic data. Commonly, an analyst working with synthetic data may be interested in how well their model can generalize to the population. By examining this aspect of synthetic data before releasing it, we can ensure that data users can derive meaningful models. In this simulation study, we fit a linear regression model of $X_2$ based on $X_1$ on a synthetic dataset and use this model to predict $X_2$ using $X_1$ from its corresponding testing dataset. Using the predicted values of $\hat{X}_2$, we obtain the RMSE using the formula $\sqrt{\frac{\sum_{i=1}^n(X_{2,i} - \hat{X}_{2,i})^2}{n}}$ and normalize it by the standard deviation of the testing $X_2$. We also replicate this process for $X_3$ with $X_1$ and $X_2$ as predictors. 

The results in Table \ref{tab: sim_specific_utility} show that our proposed methods perform reasonably well in terms of preserving valid statistical inference. Stepwise and sandwich fixed slope are our two best performers for intercept and slope differences. A possible explanation for this scenario is that we simulate data using fixed slopes and the fixed slope methods keep the slope of the median across all the quantiles and the slope estimation at the median has a higher chance of being close to the true slope. This consequently leads to smaller intercept differences, as the intercept and slope differences are correlated. In terms of RMSE, the four methods perform roughly the same. Sandwich fixed slope KNG, the best performer for general utility, has NRMSE results that are roughly 1.26\% and 4.7\% higher than the non-private quantile regression version, which is a reasonable trade-off for the differential privacy guarantee. This result shows that our proposed methods continue to maintain prediction utility even after multiple rounds of synthesis and with a small privacy loss budget of 1. Finally, we see that the extended KNG methods perform better than the pMSE mechanism across all specific utility measures. However, compared to the original KNG, the pMSE mechanism substantially improves the accuracy of regression and prediction. Overall, we find that our proposed methods offer the protection of $\epsilon$-DP with a reasonable cost in statistical validity. 

\begin{table}[H]
    \centering
    \begin{tabular}{ |c|c|c|c|c| } 
        \hline
        \diagbox{Method}{Utility} & Int Diff & Slope Diff & NRMSE $X_2$ & NRMSE $X_3$ \\ 
        \hline
        Non-private         &  2.07 (0.10) &   0.86 (0.06) &  0.317 (0.001)  & 0.193 (0.001) \\
        Stepwise - Fixed    & 14.76 (0.23) &   1.83 (0.14) &  0.326 (0.001)  & 0.201 (0.001) \\ 
        Stepwise - Varying  & 15.15 (0.12) &   5.72 (0.44) &  0.325 (0.001)  & 0.203 (0.001) \\ 
        Sandwich - Fixed    & 12.20 (0.26) &   2.05 (0.16) &  0.321 (0.001)  & 0.202 (0.001) \\
        Sandwich - Varying  & 14.95 (0.10) &   6.29 (0.44) &  0.326 (0.001)  & 0.207 (0.002) \\
        KNG                 & 70.22 (0.39) & 211.50 (1.85) &  1.705 (0.011)  & 1.601 (0.036) \\
        pMSE Mechanism      & 28.53 (1.47) &  36.40 (2.55) &  0.390 (0.007)  & 0.265 (0.006)\\
        
        \hline
    \end{tabular}
    \caption{Comparing the specific utility of synthetic simulated data generated by different methods. Means and standard errors (in parentheses) across 100 replications are reported.}
    \label{tab: sim_specific_utility}
\end{table}

\section{LBD Application} \label{app}

\subsection{Data Synthesis}
To highlight the performance and usability of the proposed methods, we apply them to SynLBD, the synthetic version of the LBD dataset.\footnote{Access to the SynLBD data can be requested through the U.S. Census \url{https://www.census.gov/programs-surveys/ces/data/public-use-data/synthetic-longitudinal-business-database.html}.} Similar to \citet{pistner2018synthetic}, we treat the SynLBD dataset as its confidential version (i.e., we label it as "raw" data) and implement KNG to generate a new DP synthetic dataset. \footnote{Code for this paper can be found on GitHub \url{https://github.com/tranntran/privatequantile}.} Table \ref{tab: synlbd_variables} shows all the variables available in the SynLBD dataset along with their definition based on the codebook \citep{synlbd_codebook}.

\begin{table}[H]
    \centering
    \begin{tabular}{ |c|p{13cm}| } 
        \hline
        \textbf{Variable}    &   \multicolumn{1}{c|}{\textbf{Definition}} \\ \hline
        sic3        &   The Standard Industrial Classification in three-digit                     form \\ \hline
        firstyear   &   The first year an establishment is observed;                             Left-censored at 1976 \\ \hline
        lastyear    &   The last year an establishment is observed;
                        Right-censored at 2000 \\ \hline
        mu          &   Multi-unit status; Coded 1 if an establishment is a                      part of a firm with two or more establishments and 0                     otherwise \\ \hline
        emp         &   Number of paid employees of an establishment as of                       March 12 annually \\ \hline
        pay         &   The reported total annual payroll in \$1,000 \\
        \hline
    \end{tabular}
    \caption{Variables contained in the current SynLBD}
    \label{tab: synlbd_variables}
\end{table}

Following the results from \citet{kinney2011towards}, establishments from each SIC code have to be modeled and synthesized separately. In this application study, we generate five DP synthetic versions, each using a privacy-loss budget of 1, for the variable employment in four industry codes. Details about each industry code are shown in Table \ref{tab: synlbd_industries}.

\begin{table}[H]
    \centering
    \begin{tabular}{ |c|c|c| } 
        \hline
        \textbf{SIC3}   &  \textbf{Industry Name}                       &   \textbf{Sample Size}    \\ \hline
        178             &  Water Well Drilling                          &   10,975                  \\ \hline
        239             &  Miscellaneous Fabricated Textile Products    &   27,396                  \\ \hline
        542             &  Meat and Fish Markets                        &   39,670                  \\ \hline
        829             &  Schools and Educational Services, NEC        &   45,709                  \\ \hline
    \end{tabular}
    \caption{Details about the four industries to synthesize}
    \label{tab: synlbd_industries}
\end{table}

We do not synthesize the first year, last year, and multi-unit status variables and treat them as publicly available data in this study.\footnote{For more information on synthesizing the first year, last year, and multi-unit status variables, see \citet{kinney2011towards}.} For the employment variable, we adapt and revise the modeling rules derived by \citet{kinney2011towards}. Specifically, to synthesize the employment of establishments founded in a specific year, we use private quantiles only. If an establishment exists during the previous year, then its employment for the next year is modeled based on employment information from the year prior. We follow the steps detailed in Section \ref{gen_syn} as our data synthesis procedure. 

For each repetition, we use the privacy-loss budget of 1 for one industry and allocate this budget equally among the employment variables over 25 years. When a year has establishments that exist the year before (termed \textit{continuers} by \citealp{kinney2011towards}) and establishments that are founded on that year (\textit{births}), we allocate 75\% and 25\% of the budget for that year on synthesizing the continuers and births, respectively. In each synthesizing model, we further divide the budget among 20 quantiles, ranging from the \nth{5} to the \nth{95} quantiles. For example, the variable employment in 1977 will have a privacy budget of 0.04. Since this year contains both continuers and births, we use a budget of 0.03 and 0.01 to synthesize continuers and births, respectively. Then, in each synthesizing model, we allocate \% to the main quantiles and \% to the median. 

Finally, we define the upper bound for all employment variables across the years as 2000. We base our decision on the statistics reported by the U.S. Bureau of Labor Statistics that 90\% of all U.S. private establishments have less than 1000 employees as of March 2021 (Appendix Figure \ref{fig:bls_establishment_size}). Like the simulation study, we top code any observations exceeding this value and set the sensitivity as 2000. We understand that this may cause some tail characteristics to be lost. However, we believe that this is a reasonable assumption for this application study.

\subsection{General Utility}
Similar to the simulation study, we use the pMSE score and KM tests to determine the similarity in distribution between the raw data and our synthetic versions. The high number of variables and the sparsity of the dataset make it harder to use the classical logistic regression model as the pMSE classifier due to convergence issues. Therefore, we propose taking a penalized regression approach and, specifically, using LASSO logistic regression model to calculate the pMSE score. For KM tests, we use the same binning approach as the simulation study, creating six bins based on the range of the data. We report the average utilities across five repetitions along with their standard error in Table \ref{tab: lbd_general_utility}.

Across all industries, \citet{pistner2018synthetic} performed best, followed by our non-private synthetic data generated with 20 quantiles. \citet{pistner2018synthetic} modeled the data at a finer granularity using 250 quantiles in their synthesis procedure. Stepwise and sandwich varying slope KNG followed those two methods. Since the SynLBD data is not generated with a fixed slope, we see a switch in performance between fixed slope and varying slope schemes.  

Overall, we observe a decrease in distributional similarity in exchange for the $\epsilon$-DP protection. Compared to the non-private method, our best performer, stepwise varying slope, has roughly ten times higher pMSE score and 6\% lower KM score. In some instances, the pMSE scores of KNG are better than stepwise varying slope and sandwich varying slope, but its KM scores are worse than those of our proposed methods. This scenario reconfirms our reasoning that there are no one-size-fits-all utility measures and that we should evaluate every synthesis method using a wide range of measures. Finally, KNG performs better than the proposed fixed slope schemes. A possible explanation is that varying slope schemes seem to model SynLBD better, and KNG's default sampling scheme is the varying slope. 

\begin{table}[H]
    \begin{adjustbox}{width=\columnwidth,center}
    \centering
    \begin{tabular}{ |c|c|c|c|c|c|c|c|c| } 
        \hline
        \multicolumn{1}{|c|}{} 
            & \multicolumn{4}{c|}{\textbf{pMSE}}
            & \multicolumn{4}{c|}{\textbf{KM}}\\
        \hline
        \diagbox{Method}{SIC} 
            & 178 & 239 & 542 & 829
            & 178 & 239 & 542 & 829 \\ 
        \hline
        \citet{pistner2018synthetic}         
            &  0.0026 & 0.0022  & 0.0021  & 0.0046
            &  947  & 915  &  929  &  912     \\
            &  (0.0002)   & (0.0005)    &  (0.0001)   & (0.0013)
            &  (1.17)   & (2.01)    &  (1.07)   & (1.37)     \\
        Non-private         
            &  0.0025 &  0.0003 &  0.0049  & 0.0056
            &  929     &  905     &  906      & 907      \\
            &  (0.0005)   & (0.0003)    &  (0.0009)   & (0.0006)
            &  (1.30)   & (1.42)    &  (1.86)   & (1.65)     \\
        Stepwise - Fixed    
            &  0.1505 &  0.1147 &  0.1452  & 0.1150 
            &  808     &  697     &  711      & 727      \\ 
            &  (0.0059)   & (0.0217)    &  (0.0030)   & (0.0122)
            &  (3.26)   & (9.48)    &  (3.39)   & (1.86)     \\
        Stepwise - Varying  
            &  0.0391 &  0.0410 &  0.0304  & 0.0501 
            &  902     &  831     &  852      & 833     \\
            &  (0.0051)   & (0.0048)    &  (0.0078)   & (0.0037)
            &  (4.79)   & (3.43)    &  (4.35)   & (4.83)     \\
        Sandwich - Fixed    
            &  0.1408 &  0.1367 &  0.1669  & 0.1093 
            &  798     &  708     &  708      & 726      \\
            &  (0.0357)   & (0.0297)    &  (0.0095)   & (0.0043)
            &  (2.43)   & (8.48)    &  (3.11)   & (1.47)     \\
        Sandwich - Varying  
            &  0.0823 &  0.0535 &  0.0573  & 0.0628 
            &  857    &  773     &  750      & 760      \\
            &  (0.0125)   & (0.0074)    &  (0.0025)   & (0.0021)
            &  (7.27)   & (10.22)    &  (4.06)   & (2.86)     \\
        KNG                 
            &  0.1141 &  0.0052 &  0.1788  & 0.0201 
            &  852     &  730     &  706      & 717      \\
            &  (0.0126)   & (0.0029)    &  (0.0006)   & (0.0091)
            &  (2.24)   & (2.64)    &  (0.00)   & (0.39)     \\
        
        \hline
    \end{tabular}
    \end{adjustbox}
    \caption{Comparing the general utility of synthetic SynLBD data generated by different methods. The mean utilities over five replications are reported with the standard error in parentheses. \citet{pistner2018synthetic} uses quantile regression with 250 quantiles to generate synthetic data, while we use only 20 quantiles in the Non-private method.}
    
    \label{tab: lbd_general_utility}
\end{table}

\subsection{Specific Utility}
\begin{figure}[H]
\begin{subfigure}{.48\textwidth}
  \centering
  \includegraphics[width=\linewidth]{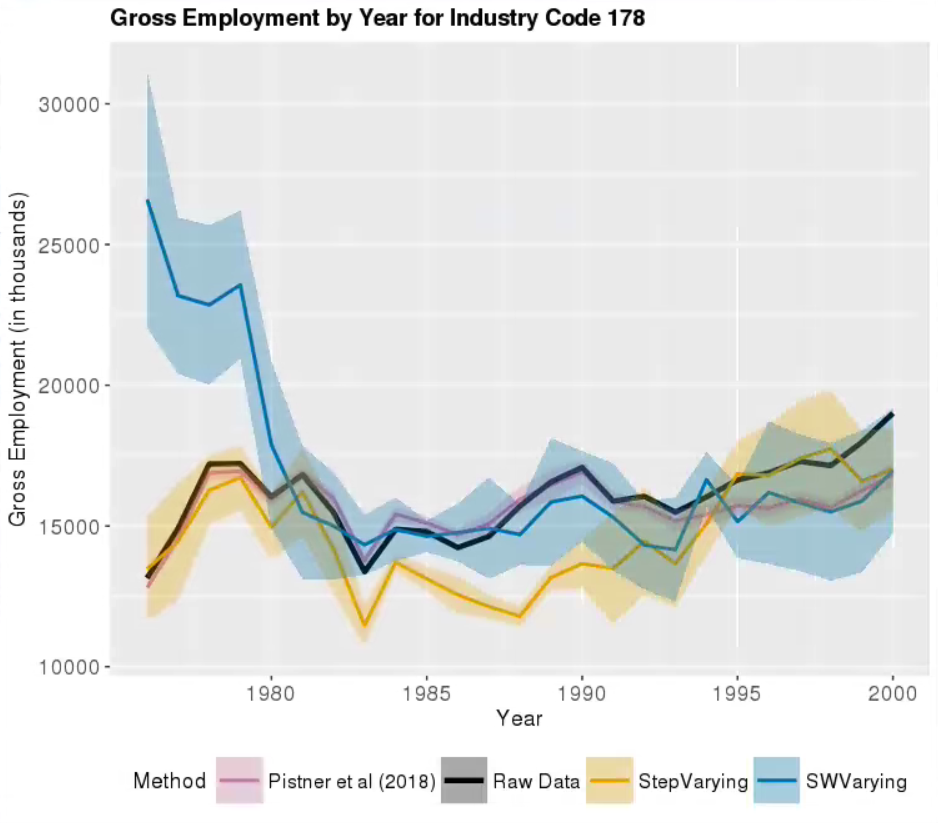}  
  \caption{Industry code 178}
  \label{fig: grossemp_sic178}
\end{subfigure}
\begin{subfigure}{.48\textwidth}
  \centering
  \includegraphics[width=\linewidth]{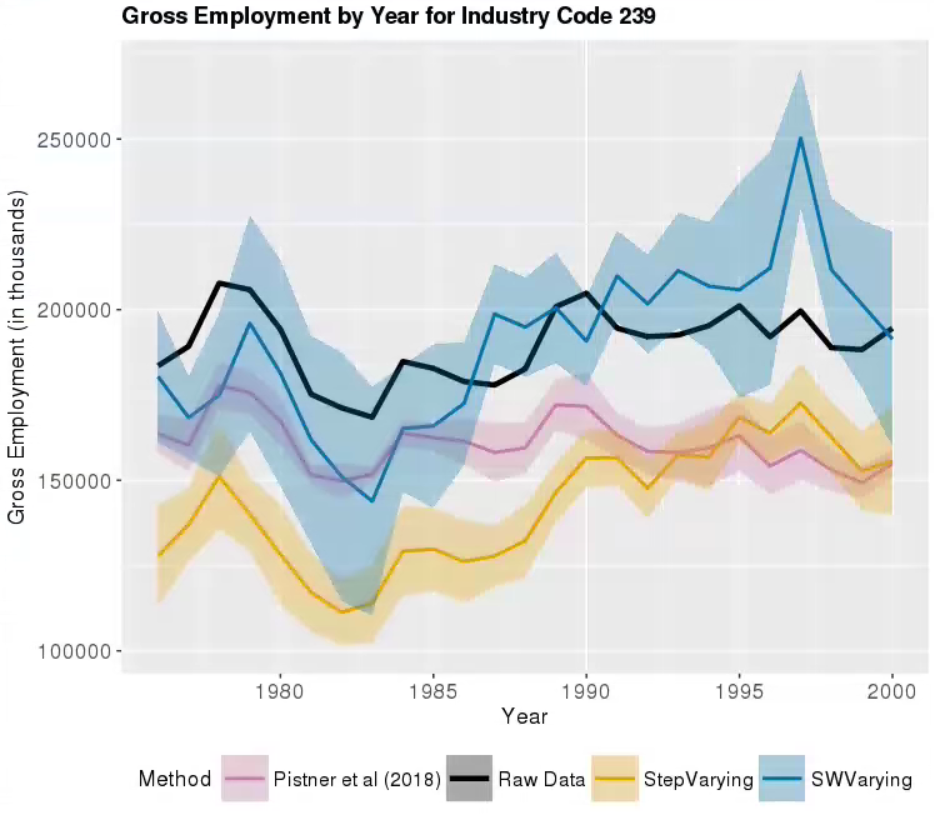}  
  \caption{Industry code 239}
  \label{fig: grossemp_sic239}
\end{subfigure}

\begin{subfigure}{.48\textwidth}
  \centering
  \includegraphics[width=\linewidth]{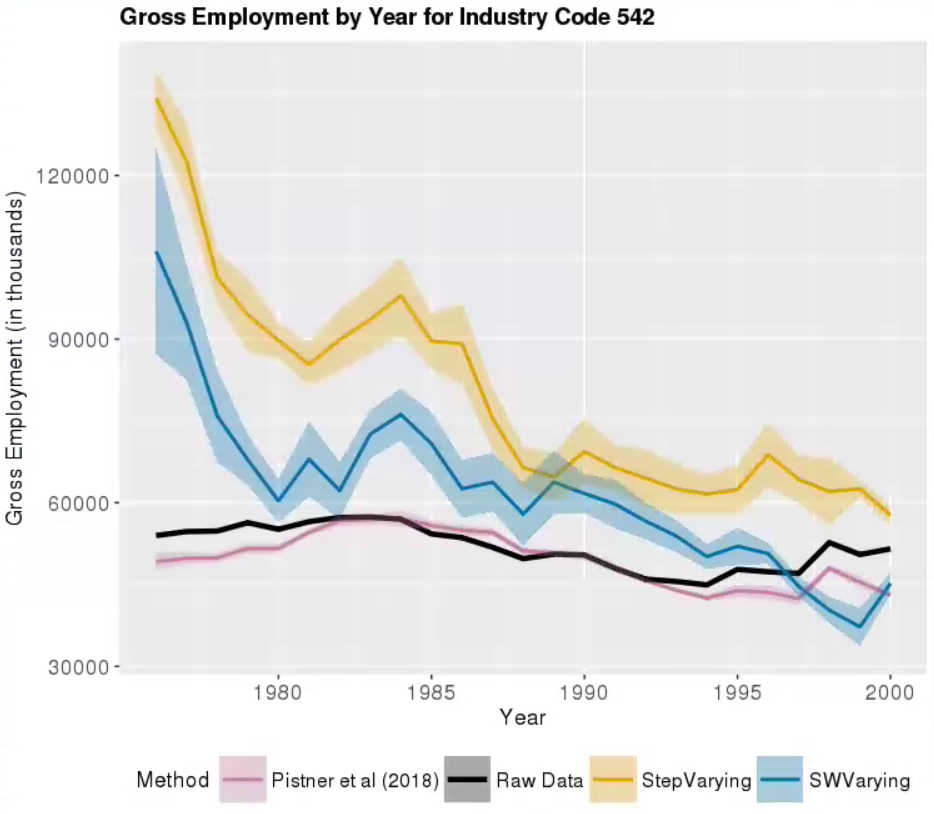}  
  \caption{Industry code 542}
  \label{fig: grossemp_sic542}
\end{subfigure}
\begin{subfigure}{.48\textwidth}
  \centering
  \includegraphics[width=\linewidth]{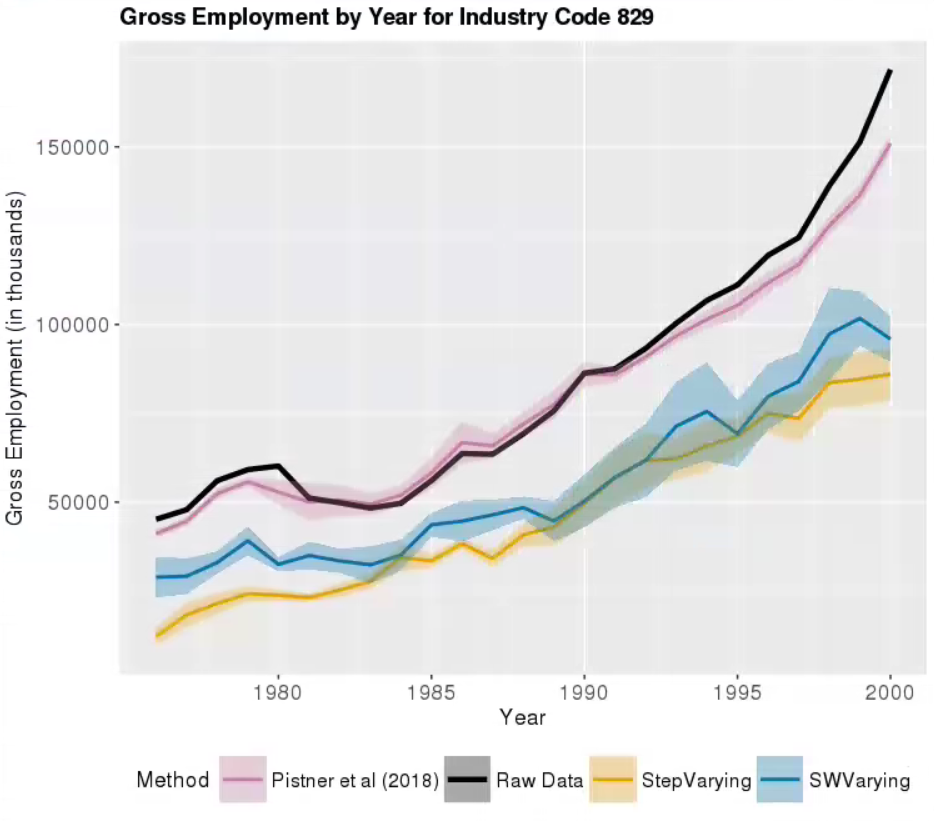}  
  \caption{Industry code 829}
  \label{fig: grossemp_sic829}
\end{subfigure}
\caption{Comparing the trend of gross employment level by year and industry. The green, red, blue, and purple colors depict the raw data, results by \citet{pistner2018synthetic}, stepwise varying slope, and sandwich fixed slope. We plot the mean across five repetitions using the straight lines and the standard deviation using the shaded area.}
\label{fig: grossemp}
\end{figure}

Since LBD plays a vital role in economic research, a good synthetic version of SynLBD needs to preserve its original trends and characteristics that researchers frequently use. Instead of using standardized coefficient difference and NRMSE to evaluate the statistical validity of our DP synthetic datasets, we follow \citet{kinney2011towards} and assess how closely they capture the actual trend of employment, job creation rate, and net job creation rate over 25 years. We plot our two best performers, stepwise and sandwich varying slope KNG, against the raw data and the non-private synthetic version by \citet{pistner2018synthetic} in Figure \ref{fig: grossemp}. The plotted synthetic results are the average of five different synthetic versions under five different seeds, and the shaded areas depict their standard deviation. Although there exist some numerical differences, our DP versions capture the trend of the raw SynLBD reasonably well, given a small privacy-loss budget of 1. Compared to \citet{pistner2018synthetic}'s synthetic data, the less favorable performance of our DP versions can be attributed to the noise added from KNG and fewer quantiles used in the synthesis procedure. 

For SIC 178, compared to the raw data, DP synthetic data generated by sandwich varying slope KNG results in a significant gap towards the start of the 25 years, but the difference is slowly narrowed down as time goes on. Stepwise varying slope KNG yields DP synthetic data that captures the movements well and results in a minor difference. In the industry with SIC 239, stepwise varying slope KNG underestimates the actual gross employment much more than sandwich varying slope KNG and the methods by \citet{pistner2018synthetic}. For SIC 542, both modified KNG methods incur big differences compared to the raw data but slowly reduce the gap towards the end of the modeling period. In the case of SIC 829, both stepwise and varying slope KNG underestimate the actual gross employment by year, but they follow the upward trend reasonably well.

\begin{figure}[H]
\begin{subfigure}{.48\textwidth}
  \centering
  \includegraphics[width=\linewidth]{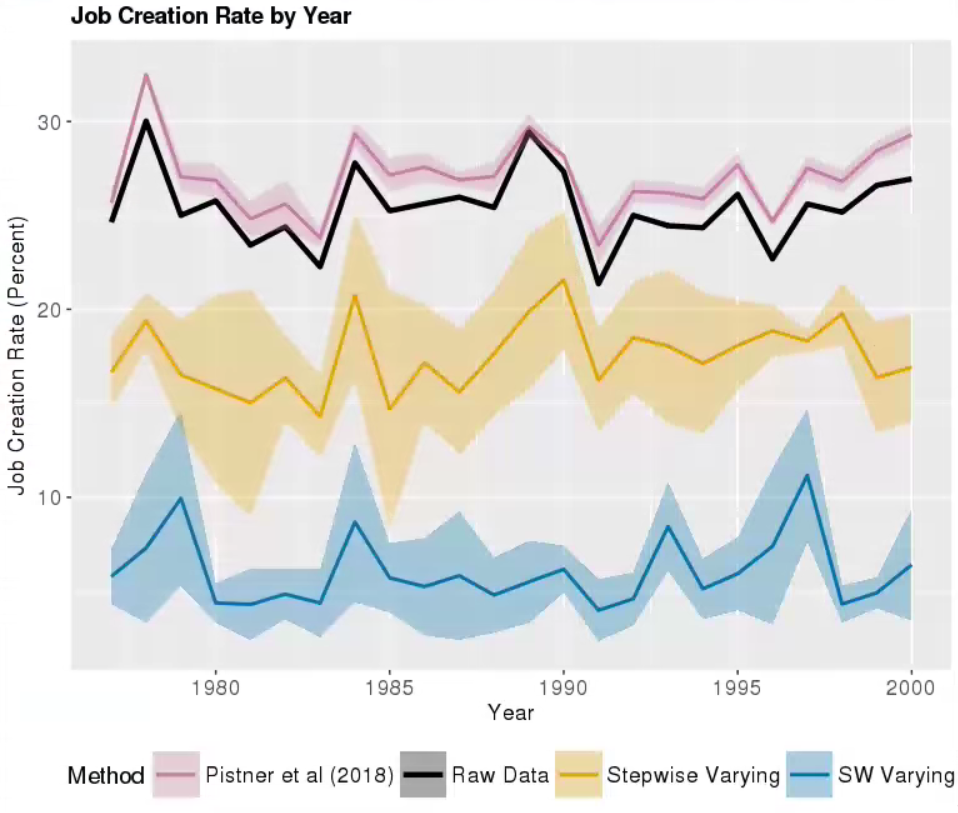}  
  \caption{Job creation}
  \label{fig:jobcreation}
\end{subfigure}
\begin{subfigure}{.48\textwidth}
  \centering
  \includegraphics[width=\linewidth]{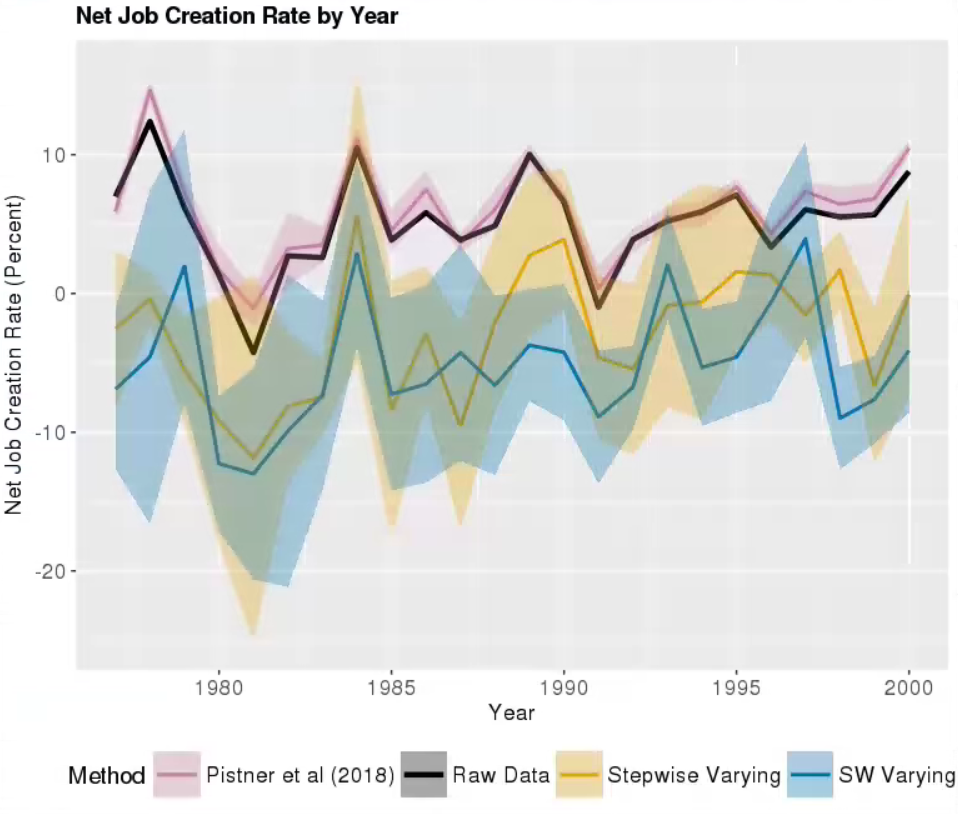}  
  \caption{Net job creation}
  \label{fig:netjobcreation}
\end{subfigure}

\caption{Comparing the trend of job creation rate and net job creation rate by year using data from four industries (178, 239, 542, 829)}
\label{fig:combinedlbd}
\end{figure}

We also aggregate results from the four industries' to get the job creation rate and net job creation rate by year (Figure \ref{fig:combinedlbd}). The job creation rate \citep{davis1998job} can be obtained using:

$$JC_t = \sum_e \frac{\max\left(0, EMP_{e,t} - EMP_{e, t-1}\right)}{Z_t},$$
where $EMP_{e, t}$ is the number of employees for establishment $e$ in year $t$ and 
$$Z_t = \sum_e Z_{e, t} = \sum_e  \frac{1}{2}\left(EMP_{e,t} + EMP_{e, t-1}\right).$$ 

The net job creation \citep{davis1998job} rate is the difference between the job creation rate and the job destruction rate, the latter of which is calculated with the following formula:

$$JD_t = \sum_e \frac{\min\left(0, EMP_{e,t} - EMP_{e, t-1}\right)}{Z_t}.$$

Overall, we observe that stepwise and sandwich varying slope KNG can reasonably retain the trend of the job creation rate over 25 years. DP  synthetic data created using stepwise varying slope is roughly 10\% lower than the raw data and \citet{pistner2018synthetic}. Meanwhile, the version created by sandwich varying slope is roughly 20\% lower on average. The difference in the net job creation rate by year is around 10\% for both methods. Although the DP synthetic versions mimic the raw data trend well, we also observe a higher variation.

\section{Discussion}
We propose implementing stepwise and sandwich KNG in the quantile regression setting for DP synthetic data generation. Quantile regression offers the flexibility and the ability to capture outliers in heavy-tailed data, while KNG provides a strong DP guarantee. Using KNG in a proposed stepwise or sandwich manner can help improve synthetic data utility and efficiency of privacy budget usage.

We evaluate our proposed methods with a simulation study and a real economic dataset, SynLBD. Simulation results show improvement in general and specific data utility compared to KNG. The utility of our DP synthetic versions is only slightly worse than the non-private synthetic data. This result shows promising progress, given we only use a privacy budget of 1. However, the standardized coefficient difference indicates the necessity of a bias-adjusting step for our DP synthetic data regression results. We are interested in finding a way to correct the bias introduced by KNG in future studies. The SynLBD application results further support the performance of our methods. As SynLBD is essential for economic research activities, maintaining the characteristics and trend of the data over time is an important indicator of whether a synthesis method is practical and usable. Stepwise and sandwich varying slope KNG can capture the trend of the raw data well despite some aggregate differences. We think the aggregate differences can be reduced with more quantiles and a higher privacy budget.

There are also a few limitations of our proposed methods. Similar to the case of quantile regression, using a large number of quantiles can increase the run time of the synthesis process. The setup of KNG also entails estimating one quantile at a time. This issue is further exacerbated if the synthesis model involves two or more variables. As the model coefficients are correlated, we have to make multivariate proposals and sample from KNG using the all-at-one Metropolis-Hastings algorithm, requiring longer run time in higher dimension cases to ensure the chain has traveled around enough. Another disadvantage of our proposed methods is that the synthetic results may be sensitive to the step size in Markov chain Monte Carlo (MCMC) and may require tuning efforts. Due to the heavy-tailed characteristics of the data, there are smaller gaps between the lower quantiles and bigger gaps between higher quantiles. Additionally, we use the currently available quantile as the starting point for the MCMC process of the next quantile to be sampled. Therefore, intuitively, we need to make proposals with appropriate step sizes to ensure that the chain can go to the correct region fast enough and converge without being influenced too much by the added noise. 

One of the ways to extend the current research is to explore other utility measures explicitly designed for DP synthetic data. In both the simulation study and the SynLBD application, we notice that a choice of the ``best" DP synthesis method varies depending on the utility measures utilized. As a result, we have to use several different utility measures to ensure that the synthetic datasets generated by our proposed methods can be assessed thoroughly. Having a standard measure for utility evaluation can help standardize our DP synthetic data comparison. Additionally, it would be interesting to find a way to account for the noise introduced by DP when performing a regression model on DP synthetic data. In the simulation study, we find that there can be a big standardized coefficient difference between the sensitive dataset and its DP synthetic versions. If there exists an approach to account for this noise instead of performing a regression analysis naively, it can help researchers using synthetic data arrive at more meaningful findings and facilitate the adoption of DP synthetic data. Finally, it would improve the efficiency of the current algorithm if a utility-maximizing function could be integrated into the algorithm to automate the MCMC step size tuning process. As previously mentioned, finding an optimal step size for KNG's MCMC sampling process is important, as the results can be sensitive to step size. This process can be time-consuming for data curators, especially when the synthesis procedure involves many variables. Having an automated system to tune and find the optimal step size can make it easier to adapt and implement KNG for synthesizing data in the DP setting.

\section*{Acknowledgements}
This research was supported in part by NSF Grant SES-1853209 and \#1702760 to The Pennsylvania State University. The creation of the Synthetic LBD was made possible through NSF Grant \#0427889. Access to the Synthetic LBD was made possible through NSF Grant \#1042181.

\newpage
\bibliography{pleasepublish}
\bibliographystyle{apalike}

\newpage
\appendix

\begin{appendices}

\section{Synthetic Data Generation Process}
\vspace*{6pt} 
\begin{algorithm}[H]
\SetAlgoLined
\KwIn{The original dataset $X = (x_1, ..., x_p)$}
\KwOut{A synthetic dataset $Z = (z_1, ..., z_p)$}
\For{each $i \in 2, ...,p$}{
    1a. Model $x_i|x_1, ..., x_{i-1}$ using quantile regression for a $m$ number of quantiles\;
    \For{each $j \in 1, ..., n$}{
        2a. Generate all possible synthetic values $y_{i, j, 1}, ..., y_{i, j, m}$ for each observation using the synthetic version of the data, $z_{1, j}, ..., z_{i-1, j}$\;
        2b. Randomly select a value corresponding to quantile $k$, $y_{i, j, k}$ and return it as $z_{i, j}$, the synthetic value of $x_{i, j}$\;
    }
}
 \caption{\citep{pistner2018synthetic} Synthesizing data using quantile regression}
 \label{alg: syndata}
\end{algorithm}
\vspace*{6pt}

\section{Simulation Study} \label{appendix_simulation}

\begin{figure}[H]
    \centering
    \includegraphics[width=\linewidth]{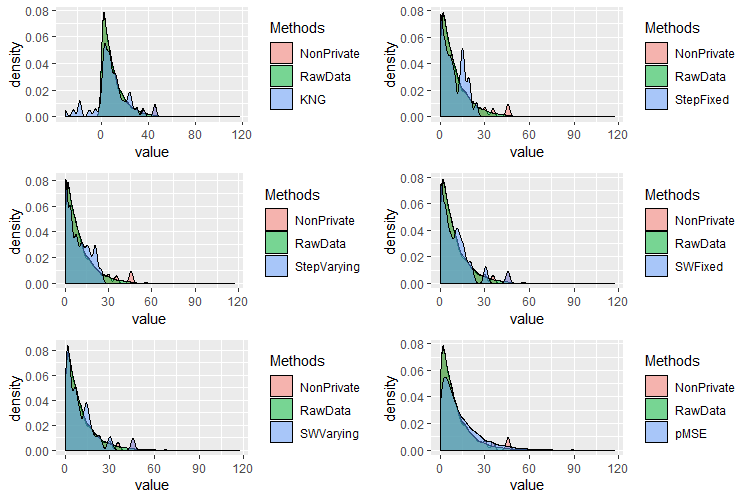}
    \caption{Comparing the distribution of synthetic $X_1$ generated by different methods}
    \label{fig:simulation_density_x1}
\end{figure}

\begin{figure}[H]
    \centering
    \includegraphics[width=\linewidth]{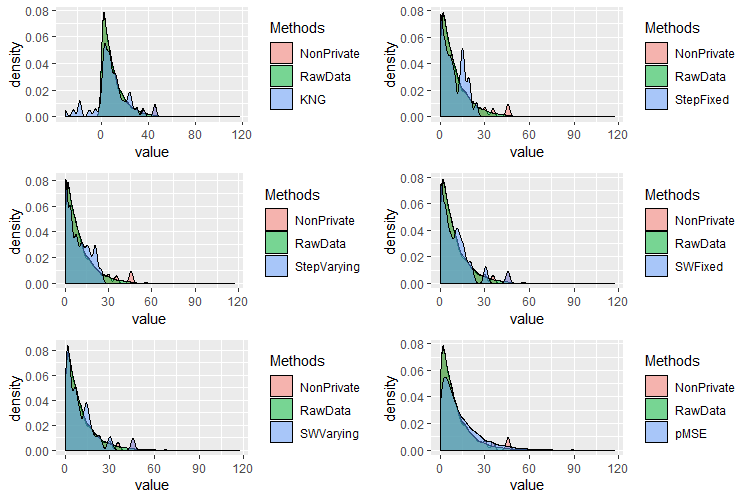}
    \caption{Comparing the distribution of synthetic $X_2$ generated by different methods}
    \label{fig:simulation_density_x2}
\end{figure}

\begin{figure}[H]
    \centering
    \includegraphics[width=\linewidth]{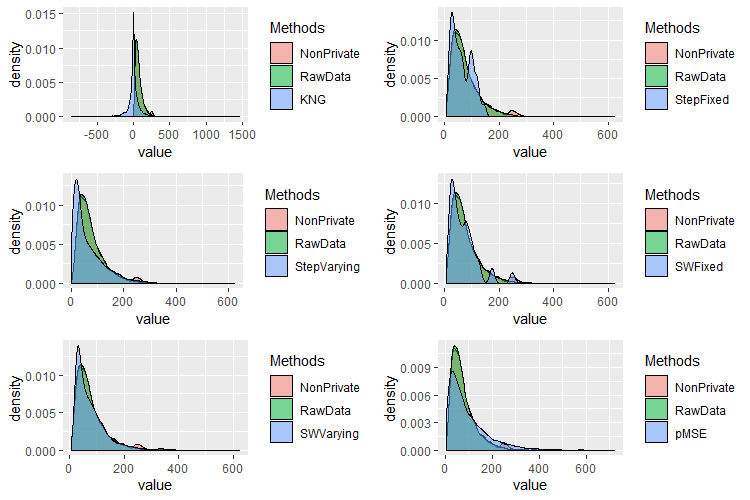}
    \caption{Comparing the distribution of synthetic $X_3$ generated by different methods}
    \label{fig:simulation_density_x3}
\end{figure}

\section{Application on the SynLBD}

\begin{figure}[H]
    \centering
    \includegraphics[width=0.75\linewidth]{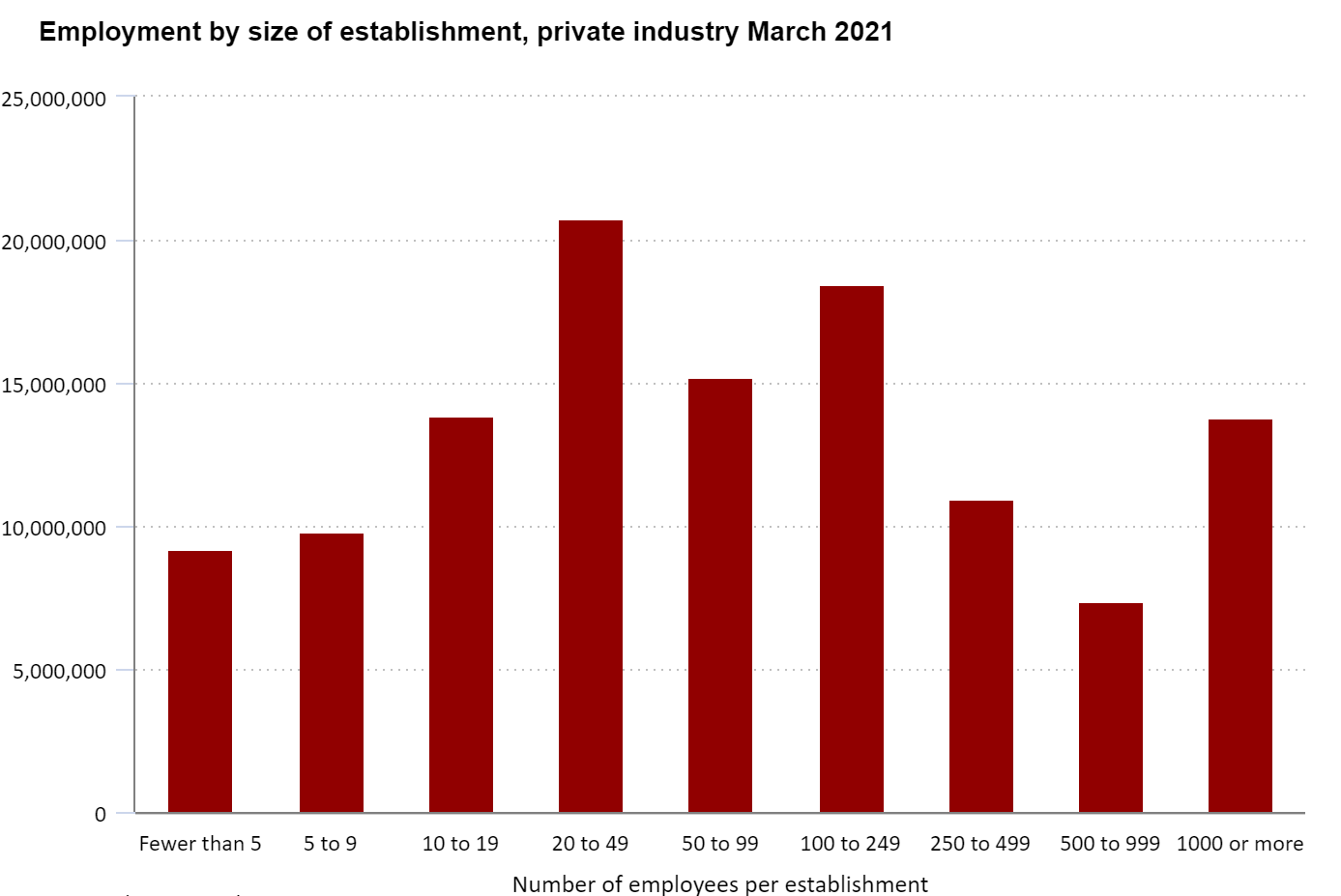}
    \caption{Employment by size of establishment in all private industry as of March 2021, reported by the U.S. Bureau of Labor Statistics }
    \label{fig:bls_establishment_size}
\end{figure}

\end{appendices}

\end{document}